\documentclass[useAMS,usenatbib]{mnras}
\usepackage[english]{babel}
\usepackage[babel]{csquotes}
\usepackage{indentfirst}
\usepackage{setspace}
\usepackage{fancyhdr}
\usepackage{graphicx}
\usepackage{booktabs}
\usepackage{amsmath}
\usepackage{amssymb}
\usepackage{mathrsfs}
\usepackage[format=plain,font=footnotesize,margin={1cm,1cm},labelfont=bf]{caption}
\usepackage{tensor}
\usepackage{placeins}
\newcommand\chandra{{\sl Chandra }}

\date{}
\pagerange{\pageref{firstpage}--\pageref{lastpage}} \pubyear{2016}
\title[X-ray spectroscopy of the $z=6.4$ quasar J1148+5251]
{X-ray spectroscopy of the $z=6.4$ quasar SDSS J1148+5251}
\author[Gallerani et al.]
{S. Gallerani$^1$\thanks{e-mail: simona.gallerani@sns.it}, L. Zappacosta$^2$, M.~C. Orofino$^1$, E. Piconcelli$^2$, C. Vignali$^{3,4}$,
\newauthor A. Ferrara$^1$, R. Maiolino$^{5,6}$, F. Fiore$^2$, R. Gilli$^4$, A. Pallottini$^{5,6,1}$,
\newauthor R. Neri$^7$, C. Feruglio$^8$
\vspace{5pt}\\
$^1$ Scuola Normale Superiore, Piazza dei Cavalieri 7, 56126, Pisa, Italy \\ 
$^2$ INAF - Osservatorio Astronomico di Roma, via di Frascati 33, I-00078, Monteporzio Catone, Italy\\
$^3$ Dipartimento di Fisica e Astronomia, Alma Mater Studiorum, Universit\'a di Bologna, viale Berti Pichat 6/2, 40127, Bologna, Italy\\
$^4$ INAF - Osservatorio Astronomico di Bologna, via Ranzani 1, 40127 Bologna, Italy\\
$^5$ Cavendish Laboratory, University of Cambridge, 19 J. J. Thomson Ave., Cambridge CB3 0HE, UK\\
$^6$ Kavli Institute for Cosmology, University of Cambridge, Madingley Road, Cambridge CB3 0HA, UK\\
$^7$ Institut de Radioastronomie Millim\'etrique (IRAM), 300 rue de la Piscine, 38406 Saint-Martin-d'H\'eres, France\\
$^8$ INAF - Osservatorio Astronomico di Trieste, via G.B. Tiepolo, 11, 34143, Trieste, Italy\\
}
\begin{document}
\maketitle
\label{firstpage}
\begin{abstract}
We present the 78-ks \chandra observations of the $z=6.4$ quasar SDSS J1148+5251. The source is clearly detected in the energy range 0.3-7 keV with 42 counts (with a significance $\gtrsim9\sigma$). The X-ray spectrum is best-fitted by a power-law with photon index $\Gamma=1.9$ absorbed by a gas column density of $\rm N_{\rm H}=2.0^{+2.0}_{-1.5}\times10^{23}\,\rm cm^{-2}$. We measure an intrinsic luminosity at 2-10 keV and 10-40 keV equal to $\sim 1.5\times 10^{45}~\rm erg~s^{-1}$, comparable with luminous local and intermediate-redshift quasar properties. Moreover, the X-ray to optical power-law slope value ($\alpha_{\rm OX}=-1.76\pm 0.14$) of J1148 is consistent with the one found in quasars with similar rest-frame 2500~\AA~luminosity ($L_{\rm 2500}\sim 10^{32}~\rm erg~s^{-1}\AA^{-1}$). Then we use \chandra data to test a physically motivated model that computes the intrinsic X-ray flux emitted by a quasar starting from the properties of the powering black hole and assuming that X-ray emission is attenuated by intervening, metal-rich ($Z\geq \rm Z_{\odot}$) molecular clouds distributed on $\sim$kpc scales in the host galaxy. Our analysis favors a black hole mass $M_{\rm BH} \sim 3\times 10^9 \rm M_\odot$ and a molecular hydrogen mass $M_{\rm H_2}\sim 2\times 10^{10} \rm M_\odot$, in good agreement with estimates obtained from previous studies. We finally discuss strengths and limits of our analysis.
\end{abstract}
\begin{keywords}
X-rays: galaxies; (galaxies:) quasars: super-massive black holes; galaxies: ISM; quasars: individual: SDSS J1148+5251 
\end{keywords}
\section{Introduction}
Studying the properties of $z\sim 6$ quasars is important to understand the formation and evolution of super massive black holes (SMBHs) at the epoch of cosmic reionization and to investigate how they interact with the host galaxies, modifying their star formation history \citep[e.g.][]{fan:2012,volonteri:2012,kormendy:2013,valiante:2016}. Hundreds of quasars have been discovered so far at $z\sim 6$ \citep[see recent numbers in][]{banados:2016}, mostly through optical telescopes, and followed-up at different wavelengths, from the near-infrared to the radio regime \citep[e.g.][]{gallerani:2017}. However, only tens of these high redshift quasars have been studied through their X-ray emission \citep{brandt:2001,mathur:2002,bechtold:2003,vignali:2003,shemmer:2006,page:2014,moretti:2014, ai:2016}. 

These studies were focused on X-ray luminous objects ($L_{\rm X}\geq 10^{44} \rm erg~s^{-1}$) and have found typical X-ray to optical power-law slopes that vary in the range $-1.9 \leq \alpha_{\rm OX}\leq -1.5$. The X-ray spectra of these sources have been modelled by a power law $N(E)\propto E^{-\Gamma}$ with $1.3 \leq \Gamma \leq 2.6$ absorbed by low column densities of gas $N_{\rm H}\leq 10^{23}~\rm cm^{-2}$ that are generally difficult to be robustly constrained.  

Local and intermediate-redshift quasars of similar X-ray luminosities are typically characterized by  $\alpha_{\rm OX}=-1.8\pm0.02$ \citep{just:2007} and $\Gamma=1.89\pm0.11$ \citep{piconcelli:2005}. Thus, $z\sim 6$ quasars do not show evident differences from their lower redshift counterparts. The only high-$z$ quasar that shows peculiar X-ray properties ($\Gamma\sim 3$ and $\alpha_{\rm OX}\sim -1.2$) is 0100+2802 at $z=6.3$ \citep{ai:2016}, powered by the most massive black hole ($M_{\rm BH}\sim 10^{10}~\rm M_{\odot}$) known so far at this redshift. However, these results are based on exploratory \chandra observations that consists of only 14 counts. In general, at these epochs, the number of X-ray observations is still sparse and often based on a limited number of counts ($\leq$ 10-30), with the exception of ULAS J1120+0641 at $z=7.1$ \citep{page:2014,moretti:2014} and SDSS J1030+0524 at $z=6.3$ \citep{farrah:2004} that have been detected with $>$100 counts (Nanni et al., submitted to A\&A).

In this work, we present X-ray observations of J1148+5251 (hereafter J1148), a quasar at $z=6.4$, obtained with the \chandra telescope. Since its discovery \citep{fan:2003,white:2003}, J1148 has been observed at several different wavelengths \citep[e.g.][]{gallerani:2008,juarez:2009,gallerani:2010,riechers:2009,maiolino:2012,gallerani:2014,cicone:2015,stefan:2015} and certainly represents the best studied case of $z>6$ quasars. In particular, from near-infrared observations of the MgII and CIV emission lines a black hole mass $M_{\rm BH} = (2 \div 6) \times 10^9 \rm M_{\odot}$ \citep{willott:2003,barth:2003} has been derived. Moreover, several CO observations have been obtained for this object in the last years \citep{riechers:2009,gallerani:2014,stefan:2015} suggesting a molecular hydrogen mass M$_{\rm H_2}\sim 2.2\times10^{10}\rm M_{\odot}$ and a size R$_{\rm H_2}\sim 2.5$~kpc \citep{bertoldi:2003,walter:2003,walter:2004}.

In this paper we analyze the X-ray properties of J1148 and discuss whether X-ray observations, can be used to further constrain the molecular hydrogen mass contained in the host galaxy of this high-redshift quasar. The paper is organized as follows: in Sec. \ref{sec_spec}, we present the \chandra\ observations, the procedure adopted for data reduction and calibration, and the results from the X-ray spectral analysis; in Sec. \ref{RDJ1148} we discuss the serendipitous, tentative detection of X-rays at the location of an optically faint $z=5.7$ quasar, previously discovered by \cite{mahabal:2005}; in Sec. \ref{sec_mod}, we describe a simple, physically motivated model for the X-ray emission from quasars and we use it to interpret our data; finally in Sec. \ref{discussion} we discuss and summarize the results of this study. 
\section{Data reduction}\label{datared}
\begin{figure}
   \begin{center}
\includegraphics[width=0.5\textwidth]{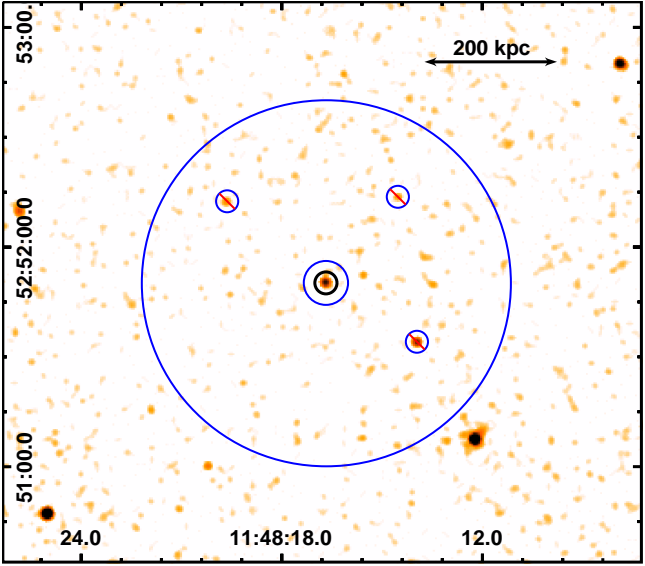}
   \end{center}
\caption{\chandra ACIS-S 0.5-7~keV exposure-corrected image centered on J1148. The image is smoothed with a Gaussian kernel of radius 0.75~arcsec (3 pixels). Thick black circular region shows the spectral extraction region for the source. Thin blue region indicates the background extraction area. Circular regions for the three point source removed from the latter are shown with barred blue circles.}
   \label{image}
\end{figure}

The 78~ks \chandra\ observation (ObsId 17127) was performed in September 2, 2015. The target was observed with ACIS-S in Very Faint mode. In order to obtain the most updated data reduction and calibration files we reprocessed the data using the script {\sl chandra\_repro} included in CIAO v.~4.8 with CALDB v.~4.7.0. We created a S3 exposure corrected 0.5-7~keV image of the field of view (see Fig.~\ref{image}) and run on it the source detection algorithm {\sl wavdetect} with scales 1.0 2.0 and 4.0 pixels and false probability threshold\footnote{This value translates in the whole S3 chip to 1 statistical fluctuation identified as point-source.} for identifying pixels belonging to a source of $10^{-6}$. The source is clearly detected with 42~background-subtracted counts with a significance of $\gtrsim 9\sigma$ (estimated within the source extraction region). The latter has been estimated by dividing the source net-counts by the Poissonian uncertainty \citep{gehrels:1986} on the background counts expected (i.e. re-normalized from the background region) in the source region. We selected  the region for source spectral extraction as a circular region of  3~arcsec radius centered on the J1148 position as reported by {\sl wavdetect}. The background extraction region  was chosen as an annular region of inner and outer radii of 6 and 50~arcsec centered on J1148 from which we have removed 3~arcsec radius circular regions centered on three nearby barely detected point-sources (see Fig.~\ref{image}). For the spectral extraction we used the script {\sl specextract} which automates spectral extraction within our selected regions and creation of relative response files.
\section{Results}
\subsection{Spectral analysis}\label{sec_spec}
For the spectral analysis we used {\sl XSPEC} v.~12.8.2 \citep{arnaud:1996}. Given the low number of counts (42) we grouped the spectrum at 1~net-count per bin and used C-stat \citep{cash:1979} with direct background subtraction \citep[Wstat in XSPEC; ][]{wachter:1979}. Furthermore, in order to check the validity of our results against possible systematics, we also performed a joint fit with the source (not background subtracted) and  background spectrum (grouped at 1~count per bin) using C-stat.  We performed all the modellings in the energy range 0.3-5~keV. Indeed the source is detected up to $\sim5$~keV. As J1148 is at $z=6.4$, we are sampling a rest-frame energy range $\sim2-37$~keV. The Galactic column density of $N_{\rm H}^{Gal}= 1.53 \times 10^{20}\,\rm cm^{-2}$, derived from \cite{kalberla:2005}, was applied to all the models. In the following, errors are quoted at $1\sigma$ level and upper limits at $90\%$.

\begin{figure*}
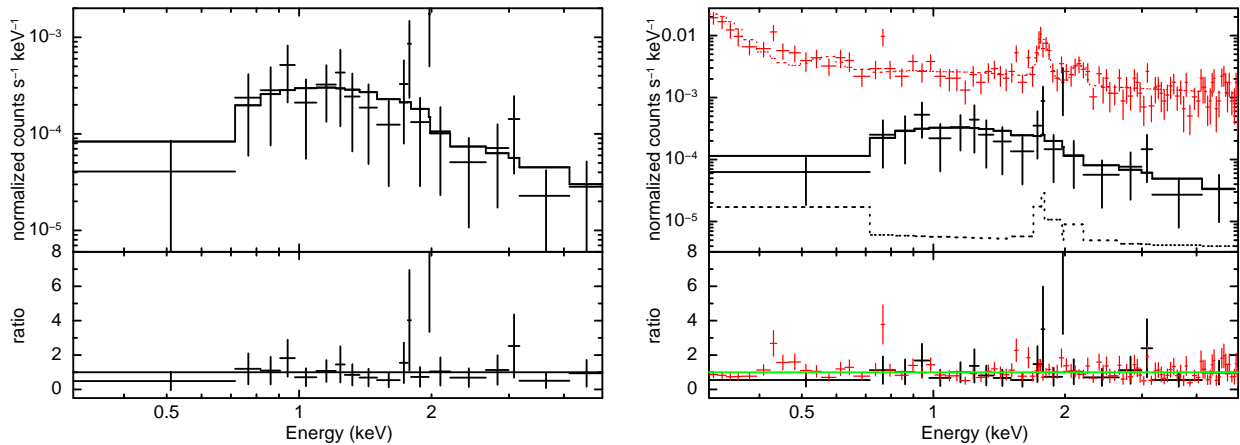

   \begin{center}
\includegraphics[width=0.33\textwidth,angle=270]{j1148_pow_spec.ps}
\includegraphics[width=0.33\textwidth,angle=270]{j1148_pow_spec_jointback_new.ps}
   \end{center}
   \caption{\chandra ACIS-S spectra and best-fit model for the simple power-law parametrization. The left panel shows the modelling on background subtracted data while the right panel reports the joint modelling on source+background (black thick) and background only (red thin). Both data and model for the latter are shown not scaled to the source+background spectral extraction area (i.e. the re-normalization of the background to the source+background spectrum has been set in the corresponding model). Background models are shown as dotted lines. The solid black line reports the total best-fit source+background spectral model.}
   \label{spectra}
\end{figure*}

We initially fitted the data with a simple power-law model (see Fig.~\ref{spectra}; left panel) obtaining a best-fit slope of $\Gamma=1.6\pm0.3$. This slope is slightly flatter than (but still consistent with) the typical  $\Gamma$ value at $\rm E>2~keV$ found for quasars \citep[$\Gamma\approx1.9$; ][]{piconcelli:2005}.   We notice that the model slightly overestimates the data in the softest X-ray portion of the spectrum. For this reason, we added  an intrinsic photoelectric component ({\tt ZWABS} in XSPEC) to investigate the possible presence of absorbing material along the line of sight to the quasar. This resulted in a rather uncertain and steeper power-law photon index, $\Gamma=2.1_{-0.9}^{+1.2}$, and a loosely constrained column density value for which we obtained an upper limit of $N_{\rm H}<1.3\times10^{24}\, \rm cm^{-2}$. In order to obtain better constraints on the column density, we fix the slope to $\Gamma=1.9$, consistently with the value found for the simple power-law model; the value is also within the range typically found for optically selected high-$z$ quasars \citep[e.g.][]{vignali:2003}. In this case we obtain a column density $N_{\rm H}=2.0^{+2.0}_{-1.5}\times10^{23}\,\rm cm^{-2}$, still consistent with no absorption at 1.3$\sigma$.  

We checked for the possible presence of the Compton reflection hump \citep{gf:1991} by adding to the simple power-law model a Compton reflection component, empirically parametrized in XSPEC by the {\tt PEXRAV} model \citep{mz:1995}. This model calculates the reflection of the power-law primary AGN flux on an infinite slab of neutral material with infinite optical depth. The reflection strength is described by the R parameter which is defined as R=$\Omega/2\pi$, where $\Omega$ defines the solid angle subtended by the reflector. We set the energy cut-off of the input power-law spectrum in {\tt PEXRAV} to 200~keV, the abundance to Solar and the inclination angle to $45~deg$. By fitting the observed spectrum with this model we obtain a photon index that is too steep ($\Gamma\approx2.8$): the reflection component dominates almost all the spectral range at the expenses of the power-law, which is relevant at the softest energies. This result is unphysical and is due to the limited statistics. Therefore we fix again the power law slope to the canonical $\Gamma=1.9$ value, obtaining a $R<0.19$. This value is in agreement with the low level of reflection  observed in higher quality spectra of lower redshift luminous broad-line AGN \citep[][Zappacosta et al. in prep.]{vignali:1999,rt:2000,page:2005}. 

We checked our results also by modelling the total source+background extracted spectrum (i.e. without any background subtraction) jointly with the background spectrum using C-stat as fit statistics similarly to what has been done in ACIS-I, for instance, by \citet{lanzuisi:2013}. In order to do this we modelled the ACIS-S3 unfocused instrumental background spectrum (grouped to 1~count per bin) as a three-segment broken power law with slopes and the two energy break free to vary and two instrumental lines at $\sim$1.77~keV and  $\sim$2.1~keV. The focused cosmic background was modelled with two thermal components with temperatures 0.07~keV and 0.2~keV with Solar metallicity \citep[parametrizing the diffuse Galactic foregrounds; ][]{lumb:2002} and a power-law for the unresolved Cosmic X-ray Background with $\Gamma=1.41$ and normalization at 1~keV set to $11.6\, {\rm photons\, cm^{-2}\, s^{-1}\, sr^{-1}\, keV^{-1}}$ \citep{dlm:2004}.\footnote{This kind of instrumental+cosmic background modeling is usually employed in faint point-source and diffuse source studies  \citep[][]{fiore:2012,humphrey:2006}} In order to get accurate global instrumental background parameters and the normalizations of the diffuse background components, we extracted a spectrum from a large 3~arcmin radius circular region (carefully excluding from it the detected point-sources) collecting $\sim28000$ counts. We modelled the background in the 0.3-10~keV energy range. 
We applied the best-fit model to the joint fit with normalizations properly scaled to the area of the source and background extraction regions. During the joint fit we tied the background normalizations with the proper area scaling and left them free to vary to adapt to the local background at the position of the source. 
For all the tested models, we obtained remarkably consistent results (see the power-law best-fit in Fig.~\ref{spectra}, right panel) further validating our spectral analysis. 

By assuming the absorbed power-law model as fiducial best-fit parametrization, we estimated fluxes in the soft (0.5-2~keV) and hard (2-10~keV) X-ray band of respectively $F_{\rm 0.5-2}=1.8\pm0.4\times10^{-15}\, \rm erg~s^{-1} cm^{-2}$ and $F_{\rm 2-10}=3.6\pm0.9\times10^{-15}\, \rm erg~s^{-1} cm^{-2}$. The intrinsic luminosities in the 2-10 and 10-40~keV energy bands are respectively of $L_{\rm 2-10}=1.4^{+0.4}_{-0.3}\times10^{45} \, \rm erg~s^{-1}$ and $L_{\rm 10-40}=1.5_{-0.3}^{+0.4}\times10^{45}  \, \rm erg~s^{-1}$. These luminosities are comparable to those typically measured in local and intermediate redshift quasars \citep[i.e.][Zappacosta et al. in prep.]{piconcelli:2005}.

Finally, we measure the X-ray to optical power-law slope defined as: $\alpha_{\rm OX}=0.3838~log (f_{\rm 2 keV}/f_{\rm 2500})$, where $f_{\rm 2 keV}$ and $f_{\rm 2500}$ are the monochromatic flux densities at rest-frame 2 keV and 2500~\AA, respectively. In particular, we use the composite quasar spectrum presented in \cite{vandenberk:2001}, to convert the broad-band z-filter measurements in $f_{\rm 2500}$, as in \cite{vignali_a:2003}. We find $\alpha_{\rm OX}=-1.76\pm 0.14$. \cite{just:2007} measured the $\alpha_{\rm OX}$ in 32 luminous quasars at $z\sim 1.5-4.5$ finding that $\alpha_{\rm OX}$ anti-correlates with the their luminosity at $2500$~\AA~($L_{\rm 2500}$). By adopting such relation \citep[see eq. 3 in][]{just:2007}, and considering that J1148 has $L_{\rm 2500}=10^{31.7}~\rm erg~s^{-1}Hz^{-1}$, we infer $\alpha_{\rm OX}=-1.73$, perfectly consistent with our estimate. This result implies that the $\alpha_{\rm OX}$ of this luminous high-$z$ quasar is comparable with the ones found in lower redshift quasars. 
\subsection{Tentative X-ray emission from a ${\small z}=5.7$ quasar}\label{RDJ1148}
The observed field of view encloses RD~J1148$+$5253, a quasar discovered by \cite{mahabal:2005} at $z=5.7$ through optical observations in the R, z, and J bands. At the position of RD~J1148$+$5253, four photons (three of these being contiguous) are detected in the observed-frame 0.5--7~keV (full) band in a circular region of 1.5\arcsec\ radius. To understand whether this might be considered a detection and derive the corresponding significance level, we adopted three methods.

At first, similarly to \cite{vignali:2001}, we extracted a 400$\times$400 pixel$^{2}$ region centered on the optical position of the quasar, excluding the immediate vicinity of the quasar itself and masking few other detected sources. This region was covered with 30,000 circles of 1.5\arcsec\ radius whose centers were randomly chosen through a Monte Carlo procedure. At this stage, to be conservative in the detection significance, we considered only the three contiguous counts (two of which in the hard, 2--7~keV, band) and found 161 and 518 extractions with at least three and two counts (out of the 30000 trials) in the full and hard band, respectively, corresponding to about 2.9$\sigma$ and 2.5$\sigma$ confidence level for a one-tailed Gaussian distribution in the two bands.

To verify this tentative detection, we considered the mean number of counts extracted in the full and hard band in the 30000 trials and assumed them as indicative of the background levels in the two bands, obtaining a probability of 2.6$\sigma$ and 2.2$\sigma$, respectively.
Finally, we ran the source-detection tool {\sc wavdetect} \cite{freeman:2002} on the full-band image to verify the results reported above. The source is detected at a false-positive threshold of 10$^{-3}$. Although {\sc wavdetect} is not properly a tool for photometry, we used its results in terms of number of counts. The source is detected with four counts, 0.7 of these due to the background. The Poisson probability of obtaining four total (source$+$background) counts when 0.7 are expected in the extraction region is $\approx5\times10^{-3}$, corresponding to $\approx3\sigma$.
From the source net counts and assuming Poissonian uncertainties \citep{gehrels:1986} we estimate the 0.5-7~keV source count-rate to be  $4.2^{+3.9}_{-2.0} \times 10^{-5} \rm counts~s^{-1}$. Assuming a power-law with canonical $\Gamma=1.9$ this corresponds to a 0.5-2~keV flux of $2.2^{+2.1}_{-1.2}\times 10^{-16} \rm erg~s^{-1}~cm^{-2}$.
We then calculate the probability that the optical source and X-ray counterparts are associated by chance. Given the 0.5-2~keV fluxes we are probing and taking as reference the source counts by \citet{lehmer:2012}, we expect to have in the ACIS-S3 field between 23 and 57 sources (accounting for $1\sigma$ uncertainty in flux). By assuming 1.5$^{\prime\prime}$ radius for the X-ray counterpart we expect a random association probability in the range $(6-16)\times10^{-4}$. 

Summarizing, without assuming any prior (e.g., the optical position of the source, as in the Monte Carlo methods), the source is tentatively detected with a significance level of 2.6--2.9$\sigma$ in the full band and 2.2--2.5$\sigma$ in the hard band. The count rate in the 0.5-7 keV band ($4.2\times10^{-5}  \rm counts~s^{-1}$) corresponds to an intrinsic luminosity of $\sim8\times 10^{43}\rm erg~s^{-1}$, assuming ($\Gamma=1.9$). Though errors are large, this luminosity is typical of sources that are intermediate between bright Seyfert and faint quasars. Moreover, given the observed J-band magnitude ($m_{\rm J}=21.45$) of RD~J1148$+$5253, and extrapolating the flux to $\lambda=2500$~\AA~with the \cite{vandenberk:2001} template ($\alpha_{\rm \lambda}=-1.5$), we infer $\alpha_{\rm OX}=-1.75$.
\section{Modelling X-ray emission in J1148}\label{sec_mod}
In Sec. \ref{sec_spec}, we have seen that if we fix the spectral slope to the value that is commonly found at lower redshift ($\Gamma=1.9$) we can marginally constrain the column density ($N_{\rm H}=2.0^{+2.0}_{-1.5}\times10^{23}\,\rm cm^{-2}$) of the absorbing gas that is intervening along the line of sight towards us. Although our detection of the X-ray absorption is at a low significance level, here we speculate on its origin and assume that it is due to metals locked in molecular clouds, distributed on $\sim$~kpc scales \citep[see the model by][hereafter G14]{gallerani:2014}. In Sec. \ref{discussion} we extensively discuss this assumption.

We compute the observed X-ray flux as $F_{\rm \nu}^{\rm obs}= F_{\rm \nu}^{\rm int} e^{-\tau}$, where $F_{\rm \nu}^{\rm int}$ is the intrinsic X-ray spectrum of J1148 and $\tau$ is the optical depth of neutral hydrogen encountered by X-ray photons in the host galaxy along the line of sight towards us. We model $F_{\rm \nu}^{\rm int}$ with a single power-law having slope $\Gamma$. Given the bolometric luminosity $L_{\rm bol}=f_{\rm Edd}L_{\rm Edd}$, where $f_{\rm Edd}$ is the Eddington ratio, and $L_{\rm Edd} = 1.26 \times 10^{38} M_{\rm BH}~\rm erg~s^{-1}~\rm M_{\odot}^{-1}$ is the Eddington luminosity, the X-ray luminosity\footnote{The subscript $X$ in quantities as $L_{\rm X}$, $F_{\rm X}$, $f_{\rm X}$ refers to the band 2-10 keV, if not differently stated.} is estimated adopting bolometric corrections $L_{\rm X} = L_{\rm bol}/f_{\rm X}$. The optical depth is given by $\tau = N_{\rm H} (1.2 \sigma_{\rm T} + \sigma_{\rm ph})$ \citep{yaqoob:1997}, where $N_{\rm H}$ is the hydrogen column density, $\sigma_{\rm T}$ is the Thomson cross-section and $\sigma_{\rm ph}$ is the interstellar photoelectric cross-section by \cite{mmcc:1983} that depends on the gas metallicity $Z$. We assume that the intrinsic spectrum attenuation is due to molecular clouds characterized by a radius $r_{\rm cl}$ and a density $n_{\rm cl}$, randomly distributed within a sphere of radius $R_{\rm H_2}$. We can thus write the column density $N_{\rm H}$ as follows:
\begin{equation}
N_{\rm H} = 2 \mathscr{N}_{\rm cl}^{\parallel} \; N_{\rm cl}, 
\label{eq:NH}
\end{equation}
where $N_{\rm cl}$ is the cloud column density given by $N_{\rm cl}=\frac{4}{\pi}r_{\rm cl}n_{\rm cl}$, and $\mathscr{N}_{\rm cl}^{\parallel}$ is the number of clouds along a line of sight:
\begin{equation}
\mathscr{N}_{\rm cl}^{\parallel}= \frac{\mathscr{N}_{\rm cl}^{tot}}{2}\frac{1}{V_{\rm H_2}}\sigma_{\rm c}R_{\rm H_2},
\end{equation}
where $\mathscr{N}_{\rm cl}^{tot}=M_{\rm H_2}/M_{\rm c}$ is total number of clouds of mass $M_{\rm c}$, $V_{\rm H_2}$ is the volume occupied by $H_{\rm 2}$, and $\sigma_{\rm c}$ is the geometrical cross section of each cloud ($\pi r_{\rm cl}^2$). Thus, $N_{\rm H}$ can be rewritten as:
\begin{equation}
N_{\rm H} = 5\times 10^{22}\left(\frac{M_{\rm H_2}}{10^{10} \rm M_{\odot}}\right)\left(\frac{R_{\rm H_2}}{2.5~{\rm kpc}}\right)^{-2}.
\end{equation}
To summarize, under the aforementioned assumptions, $F_{\rm \nu}^{\rm obs}$ is completely specified by $f_{\rm Edd}$, $M_{\rm BH}$, $\Gamma$ and $f_{\rm X}$ for the intrinsic spectrum, and $M_{\rm H_2}$, $R_{\rm H_2}$ and $Z$ for the absorption model. A comprehensive analysis of the full parameter space is hampered by the limited number of counts of the \chandra spectrum combined with the large number of free parameters of our model. Moreover, some of the free parameters are degenerate, as $M_{\rm BH}$ with $f_{\rm Edd}$ and $f_{\rm X}$. Nevertheless, J1148 represents the best studied case of $z\sim 6$ quasars, thus providing a perfect laboratory to test the hypothesis that the X-ray photons absorbing material may be constituted by molecular clouds distributed on $\sim$kpc scales. In what follows, we consider $M_{\rm BH}$ and $M_{\rm H_2}$ as free parameters of the model, and we assume for the others reasonable values taken from the literature. 

We assume that J1148 shines at the Eddington luminosity, i.e. $f_{\rm Edd}=1$. This assumption is supported by several works, e.g. \cite{willott:2003} and \cite{schneider:2015} for the specific case of J1148 and \cite{wu:2015} for the general case of high-z quasars (see their Fig. 4). For the intrinsic spectrum, we adopt $\Gamma = 1.9$ (as assumed in our analysis in Sec. \ref{sec_spec}), and the bolometric correction $f_{\rm X}=230^{+170}_{-100}$ by \cite{lusso:2012}. We note that the $f_{\rm X}$ value at the J1148 bolometric luminosity $L_{\rm bol}\sim 10^{14}~\rm L_{\odot}$ is extrapolated from results obtained at lower $L_{\rm bol}$; however, it is consistent within $\sim$1$\sigma$ with the work by \cite{hopkins:2007} (i.e. $f_{\rm X}=150 \pm 50$ ), and with the results by \cite{marconi:2004} (i.e. $f_{\rm X}=114$) for the same bolometric luminosity. More importantly, our choice is justified by the fact that if we take into account the black hole mass measured by \cite{willott:2003} and \cite{barth:2003}, $M_{\rm BH}\sim 4\times 10^9~\rm M_{\odot}$, the bolometric correction by \cite{lusso:2012} returns an X-ray luminosity ($10^{45}\leq L_{\rm X}/\rm erg~s^{-1}\leq 4\times 10^{45}$) that is perfectly consistent with our measurements (see Sec. \ref{sec_spec}).

For what concerns the absorption model, high angular resolution (0.15'', namely $<$ 1 kpc at $z=6.4$) VLA observations of the CO(3-2) emission by \cite{walter:2004} have shown that the molecular gas is extended out to radii of $R_{\rm H_2} = 2.5 \pm 1.1\rm \, kpc$. We adopt the metallicity value $Z=7\pm 3~\rm Z_{\odot}$ obtained from near-infrared observations of $3.9<z<6.4$ quasars \citep{juarez:2009,nagao:2006}, scaled for the solar metallicity $\rm Z_{\odot} = 0.0122$ \citep{asplund:2009}. 

\begin{figure}
\centering
\includegraphics[width=0.49\textwidth]{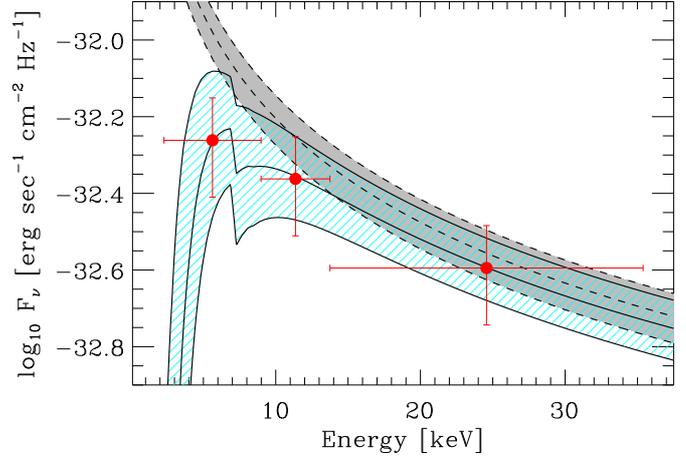}
\caption{X-ray emission model compared with the observed X-ray spectrum in the source rest frame energy range. Red circles denote data re-binned in such a way that each bin contains at least 12-14 background subtracted counts; the gray shaded region shows the intrinsic X-ray spectrum predicted for an Eddington-luminous quasar with mass of $M_{\rm BH} = 2.7\pm 0.4\times 10^9 \rm M_{\odot}$; the cyan hatched region represents our best fit model that considers absorption from an hydrogen column density $N_{\rm H} = 7.8\pm 3.0 \times 10^{22} \rm cm^{-2}$ due to a molecular hydrogen mass of $M_{\rm H_2} = 1.6\pm 0.7\times 10^{10} \rm M_{\odot}$.
}
\label{fig:model}
\end{figure}
We constrain $M_{\rm BH}$ and $M_{\rm H_2}$ by comparing our model with the X-ray data, shown in Fig. \ref{fig:model} as red filled circles. For this analysis, we re-bin the spectrum (see Fig. 2, left panel) in such a way that each bin contains at least 12-14 background-subtracted counts\footnote{The $\chi^2$ analysis can provide biased results when $n$ falls below 10-20 counts per bin \cite[][]{cash:1979,humphrey:2009}. 
Nevertheless, the large uncertainties on the adopted fiducial parameters cover a range of values larger than a possible systematic bias.}.\\ 
The flux density obtained with the fiducial unabsorbed spectrum is shown in Fig. \ref{fig:model} with the black dashed line and the gray shaded region that reflects the $M_{\rm BH}$ uncertainties; the solid line and the cyan hatched region represent the best-fit model to the observed spectrum with the corresponding $M_{\rm H_2}$ uncertainties. The values that minimize the $\chi^2$ are $M_{\rm BH} = (2.7\pm 0.4) \times 10^9 \rm M_\odot$ and $M_{\rm H_2} = (1.6 \pm 0.7) \times 10^{10} \rm M_\odot$, where the errors provide the $1 \, \sigma$ uncertainties for a chi-square distribution with one degree of freedom. The $\rm H_2$ mass range corresponds to a column density $N_{\rm H} = (0.8\pm 0.3) \times 10^{23} \rm cm^{-2}$. Resulting values of $M_{\rm BH}$ and $M_{\rm H_2}$ are perfectly consistent with previous black hole mass estimates by \cite{willott:2003} and \cite{barth:2003} and the $H_2$ mass inferred by \cite{walter:2003} from CO(3-2) observations. 
\begin{table}
\centering
\begin{tabular}{l|c|c|c}
\hline
& $M_{\rm BH}$& $M_{\rm H_2}$& $N_{\rm H}$ \\
& $[10^9 \rm M_\odot]$& $[10^{10} \rm M_\odot]$ & $[10^{23} {\rm cm^{-2}}]$ \\
\hline
Fiducial model &$2.7\pm0.4$&$1.6\pm0.7$&$0.8\pm0.3$\\
\hline
$\Gamma =   3.3$& $29 \pm 5$&	 $6.4 \pm 0.7 $ & $3.2 \pm 0.3$   \\
$\Gamma =  1.7$& $1.9 \pm 0.3 $ & 	$0.8\pm 0.6$&$0.4\pm0.3$\\
\hline			
$f_X = 130$& $1.5 \pm 0.2$& $1.5 \pm 0.6 $ & $0.7 \pm  0.3$ \\
$f_X =  400$& $4.7 \pm 0.7$& $1.6 \pm 0.6 $ & $0.8 \pm  0.3$ \\
\hline
$Z =  \rm Z_\odot$& $3.5 \pm 0.5$   & $8 \pm 3 $ & $4 \pm  1$ \\
$Z = 10~\rm Z_\odot$& $2.5 \pm 0.4$   & $1.1 \pm 0.4 $ & $0.5 \pm  0.2$ \\
\hline
$R_{\rm H_2} = 1.4~\rm kpc$  & $2.8 \pm 0.4$   & $0.5 \pm 0.1 $ & $0.9 \pm 0.3 $ \\
$R_{\rm H_2} = 3.6~\rm kpc$  & $2.7 \pm 0.4$   & $3.2 \pm 1.6 $ & $0.8 \pm  0.4$ \\	
\hline
\end{tabular}
\caption{Best values of the black hole mass $M_{\rm BH}$, molecular hydrogen mass $M_{\rm H_2}$ and hydrogen column density $N_{\rm H}$ obtained for the fiducial model ($\Gamma=1.9$, $f_{\rm X}=230$, $Z=7~\rm Z_{\odot}$, $R_{\rm H_2}=2.5~\rm kpc$), and by varying the fiducial parameters.}
\label{tab:models}
\end{table}	

Finally, we explore how the best-fit value of $M_{\rm BH}$ and $M_{\rm H_2}$ would change if we vary the fiducial parameters ($\Gamma$, $f_X$, $Z$, $R_{\rm H_2}$). We do not consider variations on $f_{\rm Edd}$ since the results would be degenerate with $f_X$. Moreover, for what concerns $Z$, we note that the measurements by \cite{nagao:2006} refers to the broad line regions (BLRs), small nuclear regions ($\leq$ few pc) that may be characterized by larger metallicity values with respect to the one in the host galaxy \citep[e.g.][]{valiante:2011}. Thus, we consider $Z=\rm Z_{\odot}$ as lower limit for the metallicity of J1148. The results of this analysis are reported in Table \ref{tab:models}. Both $M_{\rm BH}$ and $M_{\rm H_2}$ shows the largest variations (up to $\sim$ a factor of 10) with $\Gamma$. 

These results represent an important consistency check of our model. In other words, we are saying that \chandra observations can be explained by assuming that the intrinsic X-ray spectrum of J1148 is attenuated by intervening, metal-rich ($Z\geq \rm Z_{\odot}$) molecular clouds of total mass $M_{\rm H_2}\sim 2\times 10^{10}~\rm M_{\odot}$, distributed at $\sim$kpc distance from the center of the galaxy, characterized by $r_{\rm cl}\sim 1-10~\rm pc$ and density $n_{\rm cl}=10^3-10^4~\rm cm^{-3}$. 

However, given the large uncertainties that are plaguing our analysis and the novelty of our approach some caution is needed. In the next section we thoroughly discuss both strengths and limits of this study.
\section{Discussion}\label{discussion}
Estimates of $M_{\rm H_2}$ rely on the relation $M_{\rm H_2}=\alpha_{\rm CO}\times L^{'}_{\rm CO(1-0)}$, where $\alpha_{\rm CO}$ is the conversion factor, in units of $\rm M_{\odot}~(K~km~s^{-1}pc^2)^{-1}$, and $L^{'}_{\rm CO(1-0)}$ is the CO(1-0) luminosity, in units of $\rm K~km~s^{-1}~pc^2$. For what concerns the conversion factor, in Galactic molecular clouds $\alpha_{\rm CO}=0.8$, while ultra-luminous infrared galaxies (ULIRGs, $L_{\rm FIR}> 10^{12}\rm L_{\odot})$ and nuclear star-burst galaxies have $\alpha_{\rm CO}=4$. However, this factor remains highly uncertain \citep[see][for a review on this subject]{bolatto:2013} and generally presents strong variations with the metallicity \citep[e.g.][]{narayanan:2012} and with the star formation rate \citep[e.g.][]{clark:2015}. Moreover, CO(1-0) data are generally poorly constrained at high-$z$ since the CO(1-0) frequency ($\nu_{\rm RF}=115\rm~ GHz$) falls outside the ALMA bands for $z>0.4$. Thus, no $z\sim 6$ quasar has been detected so far in this transition. In the J1148 case, for example, the $M_{\rm H_2}$ estimate is based on the assumption of a constant brightness in the CO(3-2) and CO(1-0) transitions.\\ 
On the one hand, all these caveats make the possibility of measuring $M_{\rm H_2}$ through alternative, independent methods extremely appealing. On the other hand, our result is valid under the assumption that the intervening gas absorbing X-ray photons is constituted by molecular clouds (MC) distributed on $\sim$kpc scales, larger than those generally considered in more standard scenarios. \\
According to the {\it unified-model} \citep[e.g.][]{antonucci:1993,urry:1995}, the large variety of AGN types is just the result of varying orientation relative to the line of sight. In this scenario, objects that appear as optical TypeII AGN are observed along a line of sight covered by an axisymmetric dusty structure (the ``torus'') with dimensions $r_{\rm torus}\sim 0.1-10$~pc. TypeI AGNs are instead not obscured since the line of sight does not intercept the torus itself.
Since its original formulation, the {\it unified-model} has been often revisited \citep[see][for a recent review discussing the origin and properties of the central obscurer and the connection with its surrounding]{netzer:2015}. For example, another possible origin for X-ray absorption concerns with gas that is located on broad line region \cite[BLR, e.g.][]{peterson:2006} scales ($r_{\rm BLR}\sim 0.01-1$~pc) and constituted by high-density ($n_{\rm BLR}>10^9~\rm cm^{-3}$), metal-rich ($Z\sim 1-10~\rm Z_{\odot}$) clouds \citep{risaliti:2002}. This scenario (hereafter called {\it BLR-scenario}) nicely accounts for the X-ray emission variability of AGN\footnote{Variability studies of local AGN ($z<0.1$) indicate that more luminous sources vary with a lower amplitude, implying that the X-ray variability of J1148 is expected to be modest \citep[e.g.][]{lawrence:1993,almaini:2000,paolillo:2004}. However, there are hints that quasars of the same X-ray luminosity are more variable at $z>2$ \cite[e.g.][]{vagnetti:2016,manners:2002}. Nevertheless, this redshift dependence is only tentative and can be explained by selection effects \cite[e.g.][]{lanzuisi:2014,shemmer:2014}. Given the high luminosity of J1148 and the uncertain redshift dependence of AGN variability, we have ignored in our analysis possible X-ray variability effects.} that occurs on very short time scales (from a fraction of a day up to years), and thus requires the absorber to be at sub-parsec distances from the X-ray source \cite[e.g.][]{cappi:2016,bianchi:2009,risaliti:2005}. It has been also proposed by \cite{matt:2000} that while Compton-thick Seyfert 2 galaxies are observed through the torus, Compton-thin/intermediate Seyfert galaxies are obscured by dust lanes in the host galaxy.

The existence of all the aforementioned scenarios demonstrates that the origin of the X-ray absorption is far from being definitively understood and must occur on a wide range of scales. This is in line with the complex picture discussed by \cite{elvis:2012} that takes into account the BLR, the torus obscuration, and the presence of outer ($0.1-1$~kpc) absorbing regions connected to the host disk and/or dust lanes \citep[see also the discussion on the torus-galaxy connection by][]{netzer:2015}. In particular, high-$z$ quasars are expected to form in massive, over-dense regions, with dark matter halo mass $M_{\rm DM}\sim 10^{12}-10^{13}~\rm M_{\odot}$ and virial radii $r_{\rm vir}\sim 50-100$~kpc, that result from of a long history of lower mass systems merging \cite[e.g.][]{valiante:2014}. More in general, since high-redshift galaxies are thought to be very compact and gas-rich systems, it is plausible that their interstellar medium may provide substantial contribution to the X-ray emission obscuration \citep[e.g.][]{bournaud:2011,juneau:2013,gilli:2014}. Thus, the assumption of having intervening material distributed {\it also} on $\sim$kpc scales sounds plausible. 

We further note that the $H_2$ mass inferred from our analysis corresponds to a column density of neutral hydrogen $N_{\rm H}\sim 10^{23}~\rm cm^{-2}$. Since we are assuming a metallicity close to (or even higher than) the solar one, if the Galactic $A_{\rm V}/N_{\rm H}\sim 10^{-21}$ value \cite[e.g.][]{guver:2009} would apply to J1148, this column density implies an $A_{\rm V}>100$. This value is much larger that the one inferred in this object from near infrared observations \citep[$A_{\rm V}\sim 1$,][]{gallerani:2010}. This discrepancy is commonly found in several optically bright quasars \cite[e.g.][]{maccacaro:1982,granato:1997,maiolino:2001}.\\ 
In the {\it BLR scenario}, high gas column densities can provide a small $A_{\rm V}$ since BLRs are confined within the dust sublimation radius \cite[$R_{\rm sub}\sim 5$~pc in the case of J1148, from eq. 1 by][]{netzer:2015} and are consequently composed of a dust-poor gas. Moreover, anomalous properties of dust grains could also explain an $A_{\rm V}/N_{\rm H}$ value smaller than the Galactic one \cite[][]{maiolino_b:2001}. 

In fact, for a given total dust mass and fixed dust composition, a grain size distribution flatter than the Galactic one\footnote{The dust grain sizes distribution can be modelled through a power-law: $dn(a)\propto a^{-q}da$, where $a$ is the grain diameter that varies in the range $a_{\rm min}<a<a_{\rm max}$. The Galactic curve in the diffuse ISM can be described by a mixture of graphite and silicate grains having sizes distributed with $q=3.5$, $a_{\rm min}=0.005~\rm \mu$m and $a_{\rm max}=0.25~\mu$m \citep{mathis:1977}.} ($q<1$) and rich of large grains ($a_{\rm max}>10~\rm \mu m$) can reduce the expected  $A_{\rm V}/N_{\rm H}$ value by a factor of $10-100$.\\
Also in the {\it MC scenario} there are lines of reasoning that can make large gas column densities consistent with small $A_{\rm V}$ value. Ionized, atomic and molecular outflows are commonly observed towards local and high-$z$ quasars \citep[e.g.][see also \citet{fabian:2012} for a review on this subject]{feruglio:2010,rupke:2011,carniani:2015,carniani:2016,feruglio:2015}. In particular, observations of the CO(1-0) emission line in local ultra-luminous galaxies have found that a non-negligible fraction (from few\% up to 30\%) of the total $H_2$ mass is located in outflows extended on $\sim$kpc scales \citep{cicone:2014}. For what concerns J1148, while observations of CO emission lines are still not sensitive enough for detecting molecular outflows, PdBI data of the [CII]~158$\rm \mu m$ line have provided the evidence for outflowing gas in this high-$z$ quasar\footnote{We note that the [CII] emission is extended on scales ($\sim 10-30$ kpc) larger than the CO emission. The origin of this extended [CII] emission is not clear, as discussed in details in \cite{cicone:2014}. [CII] emission can both arise from dense ($n\geq 10^2~\rm cm^{-3}$) photo-dissociation regions (PDRs) and more diffuse ($n\leq 10^2~\rm cm^{-3}$) neutral/ionized gas, though the dominant contribution is expected to arise from PDRs \citep[e.g.][]{vallini:2013,vallini:2015}. From Table \ref{tab:models}, we note that if we would include contribution from PDRs, namely we would consider an $R_{\rm H_2}\geq 10~\rm kpc$, we would get a total $H_2$ mass that is a factor $>10$ larger than the one inferred by CO observations. In the case more sensitive observations would detect CO emission on the [CII] scales, our model could be easily updated to acount for larger $H_2$ masses. Viceversa, if the extended [CII] emission arises from diffuse gas, its contribution to the X-ray absorbing gas would be negligible ($\sim 10\%$). For this reason, we have not taken into account possible contribution from gas extended on the [CII] emission scales.} \citep{maiolino:2012,cicone:2015}. In a recent work, \cite{ferrara:2016} have studied the physical properties of quasar outflows finding that dust grains are rapidly destroyed by sputtering on timescales $\sim 10^4$yrs. Since sputtering preferentially destroys small grains \citep{dwek:1996}, molecular clumps can only form on massive dust grains, although the efficiency of this process is expected to be very small, challenging the theoretical interpretation of massive molecular outflows observed in local galaxies \citep[e.g.][]{feruglio:2010,cicone:2014}. To summarize, the amount of dust along the line of sight intercepted by molecular outflows is expected to be negligible, and its grain size distribution is likely dominated by large grains. This helps making the column density inferred from X-ray data consistent with the $A_{\rm V}$ inferred from optical/near-infrared observations.

To conclude, our study has the important implication that X-ray observations can provide $M_{\rm H_2}$ measurements that are independent from millimeter observations of the CO emission lines. Nevertheless, given the marginal detection of the X-ray absorption resulting from our analysis and the large uncertainty on the origin of X-ray absorbers, our interpretation of the results remains speculative and need to be tested with higher quality X-ray data. Deeper \chandra observations are necessary to better constrain the X-ray power-law photon index, and thus the column density of the absorbing gas. By detecting twice the number of X-ray photons collected in this work the error on $\Gamma$ would be reduced by $\sim$30\%. Moreover, given the novelty of the approach we have used, the same analysis should be applied to other quasars for which both $M_{\rm BH}$ and $M_{\rm H_2}$ have been measured through more standard techniques (e.g. MgII, CIV and CO emission lines). 
\section{Conclusions}\label{discussion}
We have presented the X-ray observation of the quasar J1148+5251 ($z=6.4$) obtained with 78 ks of \chandra observing time. We have clearly detected the source with 42 net counts (with a significance of $\gtrsim 9\sigma$) in the observed 0.3--7 keV energy band. We have modelled the X-ray spectrum with a power-law (photon index $\Gamma=1.9$) absorbed by a gas column density of $\rm N_{\rm H}=2.0^{+2.0}_{-1.5}\times10^{23}\,\rm cm^{-2}$. We have found that the X-ray properties of J1148 are comparable to those of lower redshift quasars: (i) we have measured intrinsic luminosities of $L_{\rm 2-10}=1.4^{+0.4}_{-0.3}\times10^{45} \, \rm erg~s^{-1}$ and $L_{\rm 10-40}=1.5_{-0.3}^{+0.4}\times10^{45}  \, \rm erg~s^{-1}$, respectively; (ii) we have checked for the presence of a reflection component over the simple power-law modelling finding a low degree of reflection from circum-nuclear material; (iii) we have measured an X-ray to optical power-law slope ($\alpha_{\rm OX}=-1.76\pm 0.14$).

In the same field of J1148, at the position of RD~J1148$+$5253, the $z=5.70$ quasar discovered by \cite{mahabal:2005},  we have detected three (two) photons in the full (hard) X-ray band. We have adopted three different methods to estimate the reliability of these detections and we have found a significance level of $\sim$2.8 (2.3)$\sigma$, i.e. the source is only marginally detected by {\sl Chandra}. Deeper observations are required to confirm the detection of this source. The tentatively inferred X-ray luminosity of RD~J1148$+$5253 ($\sim8\times 10^{43}\rm erg~s^{-1}$) is typical of sources that are intermediate between the Seyfert and quasar luminosity regime. Moreover, given its observed J band magnitude ($m_J=21.45$), we estimate $\alpha_{\rm OX}=-1.75$. 

Finally, we have used \chandra observations to test a physically motivated model that computes the intrinsic X-ray flux emitted by a quasar starting from the properties of the powering black hole and assumes that X-ray emission is attenuated by intervening, metal-rich ($Z\geq \rm Z_{\odot}$) molecular clouds distributed on $\sim$kpc scales in the host galaxy. Our analysis favors a black hole mass $M_{\rm BH} \sim 3\times 10^9 \rm M_\odot$ and a molecular hydrogen mass $M_{\rm H_2}\sim 2\times 10^{10} \rm M_\odot$, in good agreement with estimates obtained from previous studies, thus providing a solid consistency check to our model. 

This work highlights the importance of the synergy between X-ray and millimeter data for studying the properties of high-$z$ quasars and their host galaxies.
\section*{Acknowledgements}
RM acknowledges support from the ERC Advanced Grant 695671
``QUENCH'' and from the Science and Technology Facilities Council (STFC).
\bibliographystyle{mnras}
\bibliography{master}
\bsp
\label{lastpage}

\end{document}